\documentclass[preprint,preprintnumbers,amsmath,amssymb]{revtex4}
\usepackage{dcolumn}
\usepackage{bm}
\usepackage{graphicx}
\usepackage{subfigure}
\usepackage{color}
\begin{document}

\title{Probing black hole microstructure with the kinetic turnover of phase transition}
\author{Ran Li$^{a,b}$}
\thanks{Equal contributions, liran@htu.edu.cn}
\author{Kun Zhang$^c$}
\thanks{Equal contributions, zhangkun@ciac.ac.cn}
\author{Jin Wang$^{b,d}$}
\thanks{Corresponding author, jin.wang.1@stonybrook.edu}

\affiliation{$^a$ School of Physics, Henan Normal University, Xinxiang 453007, China}

\affiliation{$^b$ Department of Chemistry, State University of New York at Stony Brook, Stony Brook, NY 11794-3400, USA}

\affiliation{$^c$ State Key Laboratory of Electroanalytical Chemistry, Changchun Institute of Applied Chemistry, Chinese Academy of Sciences, Changchun, China, 130022}

\affiliation{$^d$ Department of Physics and Astronomy, State University of New York at Stony Brook, Stony Brook, NY 11794-3400, USA}

\begin{abstract}
By treating black hole as the macroscopic stable state on the free energy landscape, we propose that the stochastic dynamics of the black hole phase transition can be effectively described by the Langevin equation or equivalently by the Fokker-Planck equation in phase space. We demonstrate the turnover of the kinetics for the charged anti-de Sitter black hole phase transition, which shows that the mean first passage time is linear with the friction in the high damping regime and inversely proportional to the friction in the low damping regime. The fluctuations in the kinetics are shown to be large/small in the high/low damping regime and the switching behavior from the small fluctuations to the large fluctuations takes place at the kinetic turnover point. Because the friction is a reflection of the microscopic degrees of freedom acting on the order parameter of the black hole, the turnover and the corresponding fluctuations of the phase transition kinetics can be used to probe the black hole microstructure.
\end{abstract}

\maketitle

\section{Introduction}
Due to the pioneering work of black hole radiation by Hawking \cite{Hawking:1974sw}, black hole has been studied as the novel thermal entity \cite{Bekenstein:1973ur,Bardeen:1973gs}. It is widely acknowledged that the black hole as the macroscopic state emergent from the underlying microscopic degrees of freedom \cite{Zurek:1985gd,Strominger:1996sh,Strominger:1997eq,Lunin:2002qf,Guica:2008mu,Wei:2015iwa,Wei:2019uqg,Chen:2020nyh,Benini:2019dyp} can have phase transitions \cite{Davies:1978zz,Hut:1977nra,Hawking:1982dh,Chamblin:1999hg,Wu:2000id,Kubiznak:2012wp} in analogy to the general thermodynamic system. However, the dynamics of the black hole phase transition is still an unresolved problem.

Recently, it was suggested \cite{Li:2020khm,Li:2020nsy,Li:2020spm} that the kinetics of the black hole phase transition can be studied in terms of the Fokker-Planck equation on the free energy landscape \cite{GPT,FSW,FW,JW,JWRMP}. Due to thermal fluctuations, the horizon radius fluctuates in a stochastic way. The key is to introduce the horizon radius of the fluctuating black hole as the order parameter on the free energy landscape. The formalism is in analogy to the stochastic motion of Brownian particles \cite{ZNSM}. However, these works \cite{Li:2020khm,Li:2020nsy,Li:2020spm} are only available in the overdamped limit because the friction is assumed to be large enough. It is natural to consider how the kinetics of the black hole phase transition is affected by the friction.

To address the effect of the friction, we propose that the stochastic dynamics of the black hole phase transition should be effectively governed by the Langevin equation or equivalently described by the Fokker-Planck equation in phase space \cite{Kampen}. Kramers showed that the escaping rates of a Brownian particle must have a maximum for the intermediate damping \cite{Kramers}, which is known as the Kramers turnover \cite{HTB:1990,Melnikov}. We will demonstrate the kinetic turnover of the black hole phase transition by calculating the mean first passage time (MFPT) and its fluctuations. Our results show that the kinetics has dramatically different behavior in the high/low damping regime. Because the friction is a reflection of the microscopic degrees of freedom acting on the macroscopic degrees of freedom (horizon radius) for the black hole \cite{SDG}, studying the kinetics of the black hole phase transition in the whole friction regime provide us a new way to probe the underlying microscopic interactions \cite{Wei:2015iwa,Wei:2019uqg}.

\section{Free energy landscape of the black holes}
We focus on the phase transition of the Reissner-Nordstr\"{o}m Anti-de Sitter (RNAdS) black hole \cite{Kubiznak:2012wp} by treating the cosmological constant as the thermodynamic pressure \cite{Kastor:2009wy,Dolan:2010ha,Dolan:2011xt}. In a certain temperature regime, there are three branches of the RNAdS black hole solutions, i.e. the small, the intermediate, and the large black holes, which satisfy the Einstein field equations \cite{Kubiznak:2012wp}. The large and the small black holes are locally stable due to the time-like boundary condition of the AdS spacetime \cite{Hawking:1982dh,Witten:1998qj,Witten:1998zw}, i.e. the Hawking radiation is reflected back and then absorbed by the black hole in finite time \cite{Page:2015rxa}. However, the black hole as a thermal entity fluctuates inevitably due to the thermal noise \cite{Pavon:1988in}. In analogy to the bubbles generated during the gas-liquid phase transition, the thermal noise will give rise to the fluctuating black holes, which are not necessarily the solutions to the Einstein field equations. We construct the canonical ensemble at the specific temperature $T$ by including the three branches of the RNAdS black holes as well as the fluctuating black holes \cite{Li:2020nsy}. The black holes in the ensemble are distinguished by the continuous order parameter, the horizon radius of the black hole.

The on-shell Gibbs free energy of the three branches of the RNAdS black hole calculated from the Einstein-Hilbert action can be properly rewritten as the thermodynamic relationship of $G=M-T_H S$. We define the generalized off-shell Gibbs free energy of the fluctuating black hole as \cite{York:1986it,Whiting:1988qr}
\begin{eqnarray}\label{GibbsEq}
G=M-TS=\frac{r_+}{2}\left(1+\frac{8}{3}\pi P r_+^2+\frac{Q^2}{r_+^2} \right)-\pi T r_+^2\;,
\end{eqnarray}
where $T$ is the ensemble temperature, $r_+$ is the radius of the fluctuating black hole or the order parameter, $P=-\frac{\Lambda}{8\pi}=\frac{3}{8\pi}\frac{1}{L^2}$ is an effective thermodynamic pressure with $\Lambda$ being the cosmological constant (negative value corresponding to AdS universe) and $L$ being the AdS curvature radius, and $Q$ is the charge of the black hole.

\begin{figure}
  \centering
  \includegraphics[width=8cm]{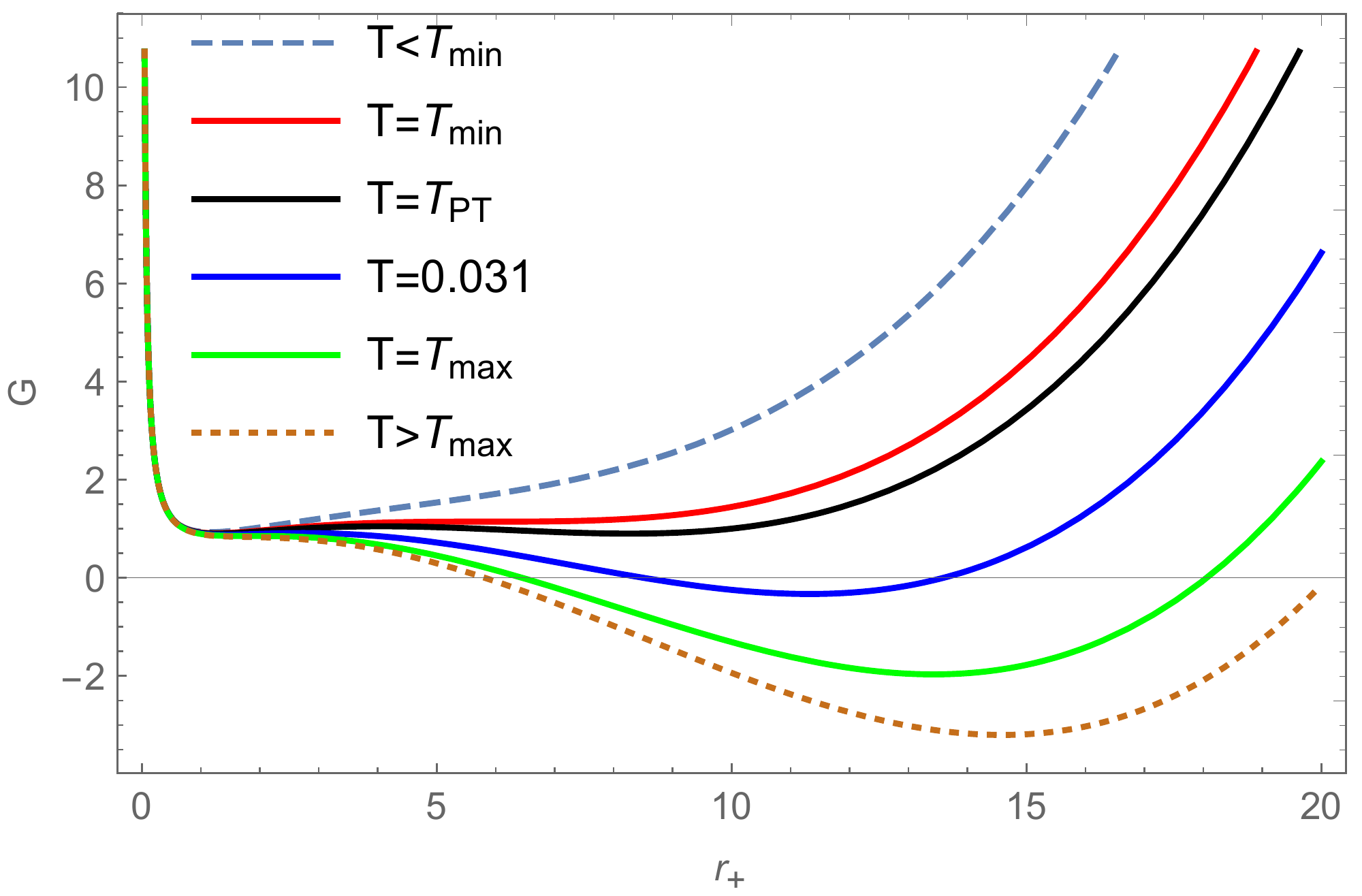}\\
  \caption{Generalized Gibbs free energy as the function of black hole radius $r_+$ for $P<P_c=\frac{1}{96\pi Q^2}$. In the plot, $P=0.32 P_c$, $Q=1$, $T_{min}=0.0256$, $T_{max}=0.0344$ and $T_{PT}=0.02703$ is the phase transition temperature. $P_c$ is the critical pressure.}
  \label{Gibbs}
\end{figure}

We quantify the free energy landscape by plotting the generalized Gibbs free energy as a function of black hole radius $r_+$  in Fig. \ref{Gibbs}. When $T_{min}<T<T_{max}$, the free energy landscape exhibits the shape of the double well. The small and the large black holes are locally or globally stable states, while the fluctuating black holes and the intermediate black hole are unstable states. These unstable black holes are treated as the transient states, which are the bridges in the phase transition process. The probability of generating a fluctuating black hole is then determined by the generalized Gibbs free energy via the Boltzmann law $p(r_+)\sim e^{-G(r_+)/k_BT}$. In this sense, the generalized Gibbs free energy can be taken as the effective potential when studying the dynamics of the phase transition. This is the free energy landscape description of the RNAdS black holes. This description is universal in studying the black hole phase transition \cite{Li:2020khm,Li:2020nsy,Li:2020spm,Wei:2020rcd,Xu:2021qyw,Wei:2021bwy}.

\section{Effective stochastic dynamics of black hole phase transition}
We discuss the effective theory of the stochastic dynamics of the black hole phase transition. For simplicity, we use the symbol $r$ to denote the black hole radius or the order parameter. Note that the generation of the fluctuating black hole is completely stochastic and the dynamical process is assumed to have the coarse-grained description by using the order parameter. Note that black hole is a macroscopic object with the number of the microscopic degrees of freedom proportional to $e^{S}$. The horizon radius as the order parameter emerges from certain combination of the microscopic degrees of freedom while the rest of the microscopic degrees of freedom plays the roll of an effective heat bath interacting with the order parameter (horizon radius). We assume that the effective stochastic dynamics describing the evolution along the order parameter is determined by three forces specified below.

The first force is the effective friction along the order parameter. It is interpreted as the interaction or the dissipation of the microscopic degrees of freedom from the effective heat bath acting on the order parameter. The friction coefficient represents the strength of this interaction. The second one is the thermodynamic driving force emergent from the interactions among all the microscopic degrees of freedom of the black hole. The free energy landscape plays the role of an effective potential. The third force represents the stochastic force that comes from the microscopic degrees of freedom of the effective heat bath on faster time scale acting on the relatively slower macroscopic order parameter. Since the black hole has a huge number of the degrees of freedom, we can characterize the statistical nature of the stochastic force by approximating its probability distribution as Gaussian. In this way, we can formulate the effective theory of the stochastic dynamics for the black hole phase transition as follows.

In analogy to the motion of a Brownian particle, we propose that the dynamics of the black hole phase transition is governed by the Langevin equation \cite{ZNSM,Kampen}
\begin{eqnarray}\label{LangevinEq}
\ddot{r}=-\zeta \dot{r}-\frac{\partial G}{\partial r}+\eta(t),
\end{eqnarray}
where the dot represents the time derivative, $\zeta$ is the friction or damping coefficient, and $\eta(t)$ is the stochastic force caused by the thermal fluctuation. The stochastic force is the Gaussian white noise with zero mean, and satisfies the fluctuation-dissipation theorem
\begin{eqnarray}\label{fluccorrelation}
\langle \eta(t)\rangle=0,\;\;\;
\langle \eta(t) \eta(s) \rangle=2 \zeta T \delta(t-s)\;,
\end{eqnarray}
where the Boltzmann constant $k_B$ is set to one. Introducing the velocity $v=\dot{r}$, the Langevin equation can be cast into the form of
\begin{eqnarray}
\dot{r}=v,\;\;\;\dot{v}=-\zeta v-\frac{\partial G}{\partial r}+\eta(t).
\end{eqnarray}
The corresponding Fokker-Planck equation in phase space is given by \cite{Kampen}
\begin{eqnarray}\label{FPEq}
\frac{\partial}{\partial t} \rho(r, v, t)&=&\left[-v\frac{\partial}{\partial r}
+\frac{\partial G}{\partial r} \frac{\partial}{\partial v}
\right.\nonumber\\&&
+\left.\zeta\left(\frac{\partial}{\partial v}v
+T\frac{\partial^2}{\partial v^2}\right)\right]\rho(r, v, t)\;.
\end{eqnarray}

Although the evolution dynamics at the emergent level is stochastic and unpredictable as described by the Langevin equation, the evolution of the distribution probability  $\rho(r, v, t)$ of the fluctuating black holes in the ensemble satisfies the linear Fokker-Planck equation and is predictable \cite{ZNSM,Kampen}.

\section{Kinetic turnover of the RNAdS black hole phase transition}
At this point, we don't know the strength of the underlying interaction or the dissipation of the microscopic degrees of freedom. However, the friction as a reflection of the underlying interactions may affect the kinetics of the black hole phase transition. In this section, we quantify the kinetics and its turnover of the RNAdS black hole phase transition. When the potential barrier on the free energy landscape is high enough, the kinetics time is amenable to the analytical derivations depending on whether $\zeta$ is large, small, or intermediate \cite{Kramers,HTB:1990,Melnikov}.

\subsection{Large $\zeta$}
For large $\zeta$, the dynamics quickly reaches the terminal velocity, so that $\dot{v}=0$. The dynamics is then effectively one dimensional in $r$. By eliminating the variable $v$, the time evolution of the reduced probability $\rho(r, t)=\int_{-\infty}^{+\infty} \rho(r, v, t) dv$ is governed by the diffusion equation \cite{Kampen}
\begin{eqnarray}\label{FPEqlargezeta}
\frac{\partial \rho(r,t)}{\partial t}=D \frac{\partial}{\partial r}\left\{
e^{-\beta G(r)}\frac{\partial}{\partial r}\left[e^{\beta G(r)}\rho(r,t)\right]
\right\}\;,
\end{eqnarray}
where $D=T/\zeta$ is the diffusion coefficient and $\beta=1/T$ is the inverse temperature. This is the equation proposed in  \cite{Li:2020khm,Li:2020nsy,Li:2020spm} to describe the phase transition dynamics of black holes.

\begin{figure}
  \centering
  \includegraphics[width=8cm]{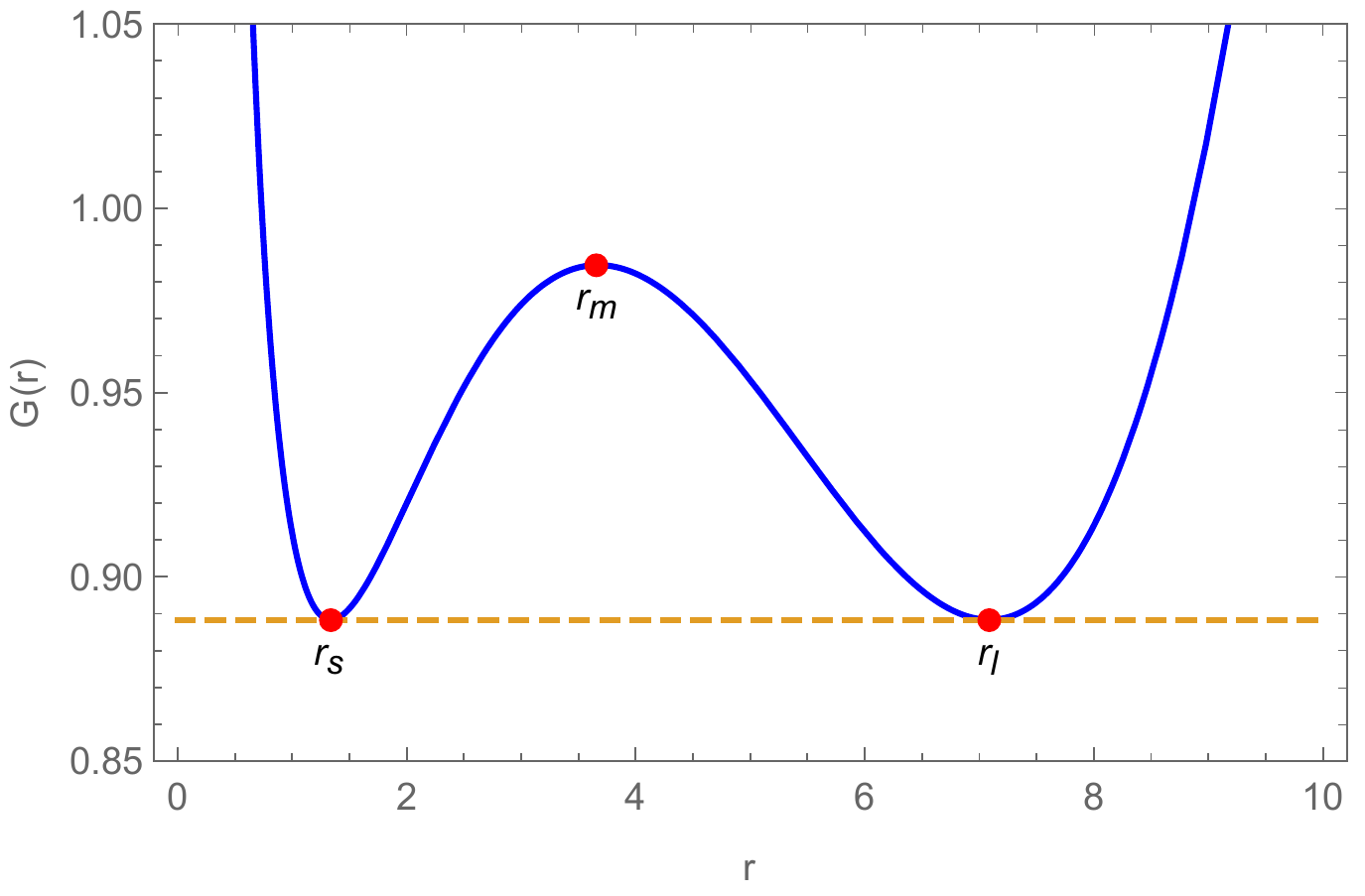}\\
  \caption{Generalized Gibbs free energy at the phase transition point. In the plot, $P=0.4 P_c$, $Q=1$, and$T=T_{PT}=0.0298$. The radii of the small, the intermediate, and the large black holes are $1.34$, $3.67$, and $7.09$. }
  \label{GPlot}
\end{figure}

Without loss of generality, we consider the MFPT of the small black hole state $(r=r_s)$ escaping over the effective potential barrier $(r=r_m)$ on the free energy landscape as depicted in Fig.\ref{GPlot}. By choosing the initial distribution $\rho(r, 0)=\delta(r-r_s)$ and the absorbing boundary condition $\rho(r_m, t)=0$ at the end, the formal solution to the diffusion equation (\ref{FPEqlargezeta}) is \cite{ZNSM}
\begin{eqnarray}
\rho(r, t)=e^{t\mathcal{D}} \delta(r-r_s)\;,
\end{eqnarray}
with the operator $\mathcal{D}=D \frac{\partial}{\partial r}\left[
e^{-\beta G(r)}\frac{\partial}{\partial r}e^{\beta G(r)}\right]$.

Define $\Sigma(t)=\int_{0}^{r_m} \rho(r, t) dr$ to be the probability that the state has not made a first passage by time $t$. Note that the first passage time is a random variable. The distribution of the first passage is given by $F_p(t)=-\frac{d\Sigma(t)}{dt}$. Then the MFPT is \cite{ZNSM}
\begin{eqnarray}\label{taverage}
\langle t \rangle=\int_{0}^{+\infty} t F_p(t) dt=\int_{0}^{+\infty} dt\left(e^{t\mathcal{D}^{\dagger}}\textrm{1}\right)\;,
\end{eqnarray}
with the adjoint operator $\mathcal{D}^{\dagger}=D e^{\beta G(r)} \frac{\partial}{\partial r}\left[
e^{-\beta G(r)}\frac{\partial}{\partial r}\right]$. The MFPT satisfies the adjoint equation $\mathcal{D}^{\dagger}  \langle t \rangle=-1$, the solution of which is given by \cite{ZNSM}
\begin{eqnarray}\label{MFPTexp1}
\langle t \rangle&=&\frac{1}{D}\int_{r_s}^{r_m}dr \int_{0}^{r}dr'  e^{\beta (G(r)-G(r'))}\;,
\end{eqnarray}
where the reflecting boundary condition at $r=0$ is imposed. One can get an approximate formula for the MFPT. The generalized Gibbs free energy near the minimum and the maximum can be approximated by the quadratic expansion as
\begin{eqnarray}
G(r)&=&G(r_{s/m})\pm \frac{1}{2}\omega_{s/m}\left(r-r_{s/m}\right)^2+\cdots\;.
\end{eqnarray}
By performing the Gaussian integral, one can obtain the approximate expression of the MFPT \cite{ZNSM}
\begin{eqnarray}\label{large_fric_formula}
\langle t \rangle\approxeq \frac{\pi\zeta}{\omega_{s}\omega_{m}}
e^{\beta W}\;,
\end{eqnarray}
where $W=G(r_m)-G(r_s)$ is the barrier height. The MFPT is proportional to the friction coefficient. We can also see clearly that the MFPT is exponentially related to the barrier height between the two states (the small and the intermediate black holes in this case). The factor in front of the exponential also contributes to the kinetics through the fluctuation $\omega_{s}$ around the basin and $\omega_{m}$ at the top of the barrier, although to a much lesser degree than that from the barrier height in the exponential.

The friction coefficient represents the coupling strength of the microscopic degrees of freedom between the effective heat bath and the order parameter. In the strong coupling regime, i.e. in the large friction regime, increasing the friction coefficient results in the slowing down of the state transitions from one black hole state to another.

\subsection{Small $\zeta$}
In the case of extremely weak friction, by introducing the action variable $I$ \cite{Kampen}
\begin{eqnarray}
I(E)=\oint p(r) dr\;, \;\;\;p(r)=\sqrt{2m(E-G(r))}\;,
\end{eqnarray}
the Fokker-Planck equation (\ref{FPEq}) can be reduced to the diffusion equation again in the energy space
\begin{eqnarray}
\frac{\partial}{\partial t} \rho(I,t)
=\zeta T \frac{\partial}{\partial I} e^{-E/T} \frac{2\pi I}{\omega(I)}
\frac{\partial}{\partial I} e^{E/T} \rho(I,t)\;,
\end{eqnarray}
where $\omega(I)=2\pi \frac{\partial E}{\partial I}$ is the angular frequency at the action $I$. In analogy to the large friction case, the MFPT for the black hole state with the initial energy $E$ to arrive at the top of the barrier is determined by the adjoint equation
\begin{eqnarray}
\zeta T e^{E/T} \frac{\partial}{\partial I} I \frac{\partial I}{\partial E}
 e^{-E/T} \frac{\partial}{\partial I}  \langle t(E) \rangle=-1\;.
\end{eqnarray}
The solution is then given by \cite{Kampen}
\begin{eqnarray}
 \langle t(E) \rangle=\frac{1}{\zeta T}\int_{E}^{W}\frac{dE'}{I'} \exp\left[\frac{E'}{T}\right]
 \int_0^{I'}\exp\left[-\frac{E''}{T}\right]dI''\;.
\end{eqnarray}
Using the parabola approximation of the generalized Gibbs free energy, one can get the MFPT \cite{Kampen}
\begin{eqnarray}\label{small_fric_formula}
\langle t \rangle =\frac{2\pi T}{\zeta \omega_s I(W)}\exp\left(\beta W\right)
\approx \frac{T}{\zeta W} \exp\left(\beta W\right)
\end{eqnarray}
The small friction coefficient is a result of the weak coupling of the microscopic degrees of freedom between the effective heat bath and the order parameter. In the weak coupling regime, the dynamics becomes ballistic in the order parameter and the effective energy or action is almost conserved. Increasing the friction can enhance the effective diffusion in energy or action space. Correspondingly, the transition rate becomes faster when increasing the friction coefficient.

The kinetic time characterized by MFPT is shown to be a linear with the friction efficient in the high damping regime and inversely proportional to the friction efficient in the low damping regime. Therefore, we have demonstrated the turnover of the kinetics of the black hole phase transition analytically. This implies the underlying interactions of the microscopic degrees of freedom between the effective heat bath and the order parameter affects the kinetics of the black hole phase transition. It is expected that this property can be used to probe the coupling strength of the interaction among the microscopic degrees of freedom of the black hole.

\section{Numerical results}
Note that the analytical formulas of the MFPT is only valid in the high potential barrier case \cite{Kramers,HTB:1990,Melnikov}, i.e. $\beta W\gg 1$. In order to reveal the kinetic turnover of the phase transition more precisely, we invoke the numerical method. We have solved the Langevin equation Eq.(\ref{LangevinEq}) to collect the trajectories of the black hole state evolution. Although the initial condition for the black hole state is chosen as the small black hole, the trajectories of the simulations are different because of the stochasticity of the thermal fluctuation. We can read off the time that the black hole used to escape from the small black hole to the intermediate black hole. This time is just the first passage time we have defined. In this way, we can obtain the MFPT as long as we have enough trajectories. In the simulations, we have collected $10000$ trajectories for each friction coefficient.

\begin{figure}
  \centering
  \includegraphics[width=8cm]{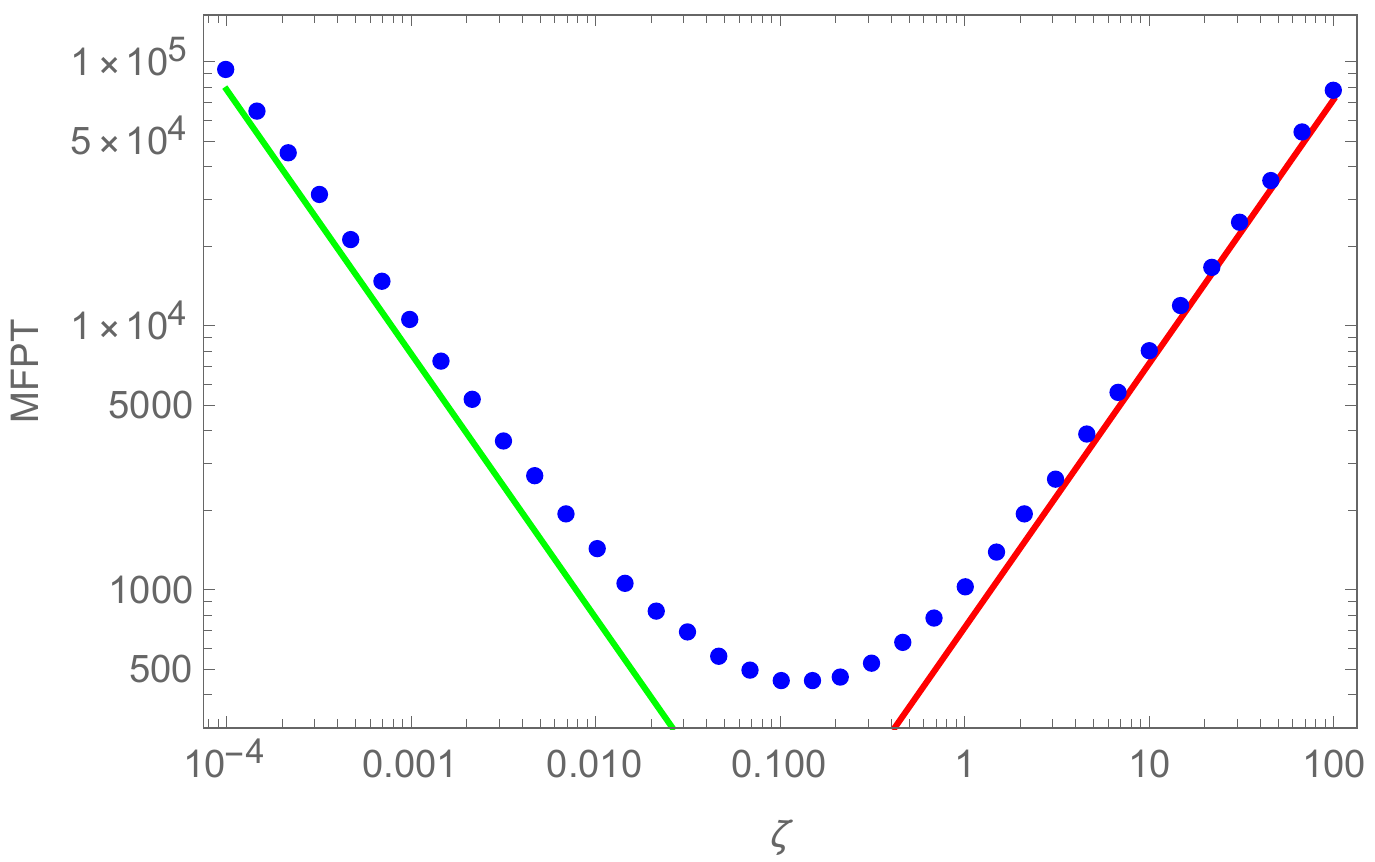}\\
  \includegraphics[width=8cm]{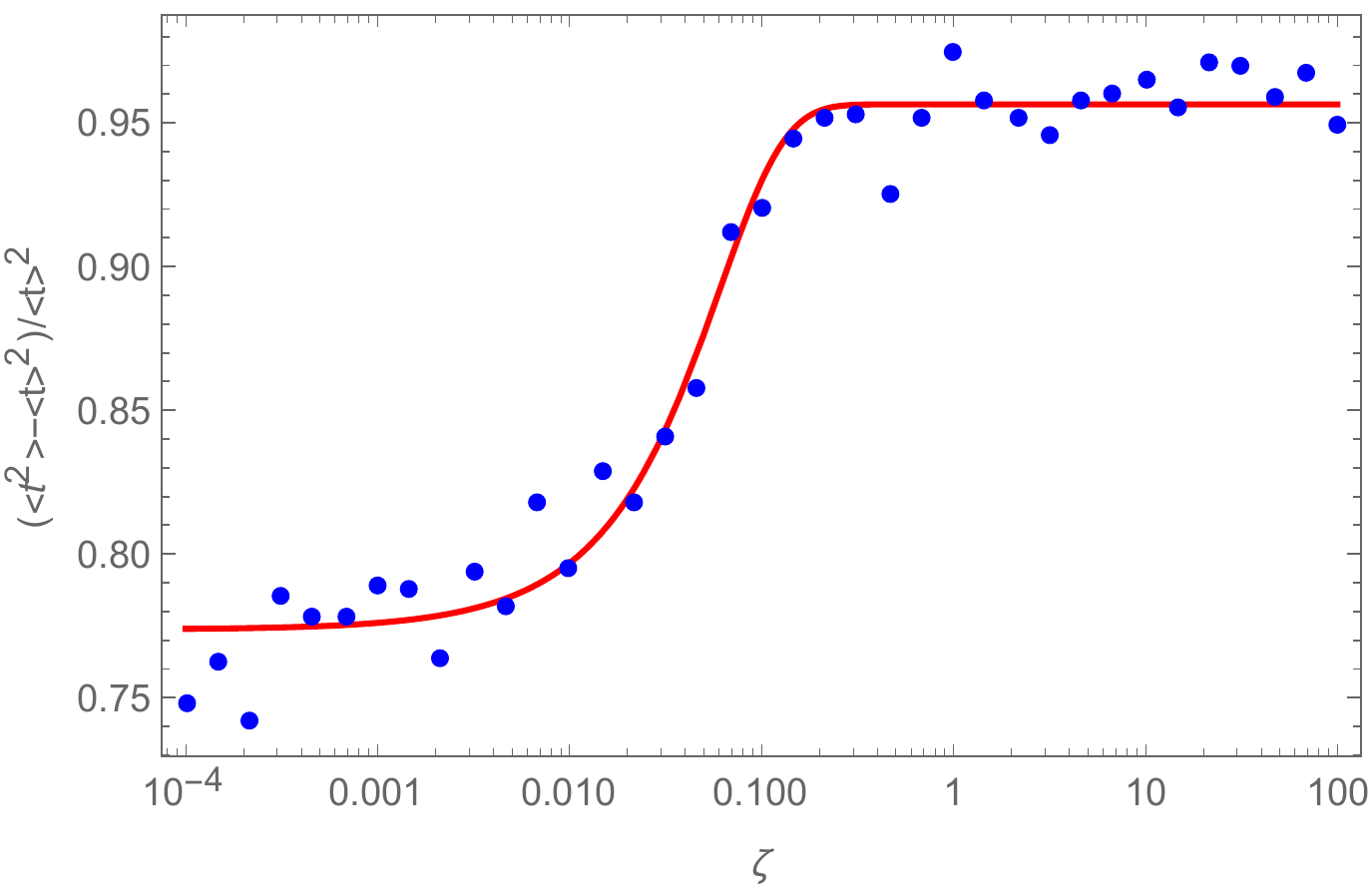}\\
  \caption{The MFPT (upper panel) and the relative fluctuations (lower panel) as the functions of the friction coefficient $\zeta$. In the upper panel, blue dotted line is the numerical results obtained by simulating the Langevin equation while red, and green lines are plotted by using the MFPT formulas for the large and the small friction coefficients. In the lower panel, the dotted line is the numerical results and the red line is the corresponding plot of the fitting function given by $a\tanh(b\zeta)+c$. In this plot, $P=0.4P_c$, $Q=1$, and $T=T_{PT}=0.0298$. The ratio of the barrier height and the temperature is $3.23$. The analytical formulas of the MFPT are validated.}
  \label{NumResult}
\end{figure}

The kinetic time characterized by MFPT and its fluctuations as the functions of the friction coefficient are plotted in Fig.\ref{NumResult}. The boundary condition used in the numerical simulation is consistent with the absorbing boundary condition $\rho(r_m, t)=0$ to derive Eq.(\ref{large_fric_formula}). It is shown that the numerical results coincide with the analytical formulas Eq.(\ref{large_fric_formula}) and Eq.(\ref{small_fric_formula}) in the small and large friction limit, i.e. the MFPT of the kinetics of the black hole phase transition is a linear function of the friction efficient in the high damping regime and inversely proportional to the friction efficient in the low damping regime. The fluctuations of the kinetics show a sharp switching between the small fluctuations in the low damping regime and the large fluctuations in the high damping regime. The transition from the small fluctuations to the large fluctuations in kinetics takes place at the turnover point of the kinetics. In addition, the fluctuations in the kinetics is monotonic with the friction. Since the strength of the stochastic force is proportional to the friction as shown in Eq.(\ref{fluccorrelation}), the larger the friction is, the bigger the strength the stochastic driving force is. The bigger stochastic force can lead to the larger fluctuations in the kinetics. These characteristics can be used to probe the strength of the friction or the dissipation. Since the friction coefficient represents how strong the order parameter interact with the microscopic degrees of freedom of the black hole, the dramatically different behavior of the kinetics and the corresponding fluctuations of the black hole phase transition at the high/low damping regime can be used to probe the coupling strength of the microscopic degrees of freedom of the black hole.

We have also shown the numerical results of the kinetics of the black hole phase transitions for different pressures (corresponding to different cosmological constants) and electric charges (details in Appendix B). The kinetics of the black hole phase transition and the associated fluctuations versus friction show similar behaviors. This suggests the universality of the kinetics for the black hole phase transition. On the other hand, due to the delicate balance between energy (or enthalpy) and entropy multiplied by temperature \ref{GibbsEq}, the resulting free energy landscape provides the thermodynamic origin for the kinetic behavior of the black hole phase transition \cite{Li:2020spm}. As the pressure (the absolute value of the cosmological constant) or the electric charge increases, the free energy barrier from the small to the large black hole through the intermediate transition state decreases. Therefore, it is easier or faster to switch from the small black hole to the large black hole when increasing the pressure (the absolute value of the cosmological constant) or the electric charge.

{\it Conclusion}
In summary, we have studied the stochastic dynamics of the phase transition of the RNAdS black holes. The dynamics of black hole phase transition at the macroscopic emergent level can be described by the evolution along the order parameter (black hole radius). We propose that the stochastic dynamics of the black hole phase transition caused by the thermal fluctuations can be effectively described by the Langevin equation or equivalently by the Fokker-Planck equation in phase space, where the generalized Gibbs free energy plays the role of the effective potential. Our framework is different from the Stochastic semi-classical gravity based on the Einstein-Langevin equation \cite{Hu:2003qn}.

Notice that the analogy between this approach to the emergent black hole dynamics and condensed matter physics such as superconductor and superfluidity. The associated order parameter such as the coherent wavefunction of copper pairing for superconductivity emerges from the microscopic degrees of freedom of the electronic interactions. The Landau-Ginzburg free energy function emerges from the microscopic degrees of freedom has been successfully applied to describe the associated dynamics. In the same token, the dynamics of the black hole phase transition at the macroscopic level can be described and determined by the forces emergent from the underlying microscopic degrees of freedom.

The effect of the friction coefficient on the kinetics of the RNAdS black hole phase transition is also addressed. The analytical and the numerical studies confirmed that the MFPT of the kinetics for the small/large RNAdS black hole phase transition is linear with the friction efficient in the high damping regime and inversely proportional to the friction efficient in the low damping regime. The fluctuation of the kinetics is shown to be large/small in the high/low damping regime. Our results show that the kinetics of the black hole phase transition has dramatically different behavior in the high/low damping regime. Because the friction is a reflection of the coupling of the microscopic degrees of freedom with the order parameter, we can conclude that studying the kinetics and the fluctuations of the black hole phase transition in the whole friction regime can provide us a new way to probe the underlying interactions of the microscopic degrees of freedom, i.e. the microscopic structure of the black holes.

Although the escaping kinetics from the small black hole state to the large black hole through the intermediate black hole state is analyzed in the present work, the main conclusions for the escaping kinetics and its fluctuations versus friction from the large black hole state to the small black hole through the intermediate black hole state will not be changed. This work provides a general framework to study the stochastic dynamics of the black hole phase transition and the black hole microstructures based on the free energy landscape. We expect that the discussion can be generalized to other types of black hole phase transition.

\section*{Appendix A: Analytical derivation of MFPT for the intermediate $\zeta$}

For completeness, we also present the derivation of the MFPT for the intermediate $\zeta$. For intermediate $\zeta$, it is convenient to find the stationary solution of Eq.(\ref{FPEq}). Around the barrier, by substituting the ansatz
\begin{eqnarray}
\rho(r,v)=\xi(r,v) \exp \left\{-\left[\frac{1}{2}v^2+G(r)\right]/T\right\}
\end{eqnarray}
into the stationary Fokker-Planck equation, we have \cite{Kampen}
\begin{eqnarray}\label{XiEq}
\left[-v\frac{\partial}{\partial r}
-\left[\omega_m^2(r-r_m)+\zeta v\right]\frac{\partial}{\partial v}
+\zeta T\frac{\partial^2}{\partial v^2}\right]\xi(r,v)=0\;,
\end{eqnarray}
where the parabola approximation to the generalized Gibbs free energy is used near the intermediate black hole. Trying a special solution
\begin{eqnarray}
\xi(r,v)=f((r-r_m)+\alpha v)=f(u)\;,
\end{eqnarray}
Eq.(\ref{XiEq}) transforms into \cite{Kampen}
\begin{eqnarray}\label{fEq}
-\left[\left(1+\zeta \alpha\right)v+\omega_m^2\alpha(r-r_m)\right]f'(u)
+\zeta T \alpha^2 f''(u)=0\;.
\end{eqnarray}
In order for the above equation to become a proper differential equation, the factor in front of the first derivative must be proportional to $u$, i.e.
\begin{eqnarray}\left[\left(1+\zeta \alpha\right)v+\omega_m^2\alpha(r-r_m)\right]=-\lambda u
\end{eqnarray}
should be satisfied, which gives
\begin{eqnarray}
\lambda_{\pm}&=&-\frac{\zeta}{2}\pm\sqrt{\omega_m^2+\frac{\zeta^2}{4}}\;,\nonumber\\
\alpha_{\pm}&=&\frac{\lambda_{\pm}}{\omega_m^2}\;.
\end{eqnarray}
In this way, Eq.(\ref{XiEq}) and (\ref{fEq}) can be easily solved. Near the bottom of the small black hole state well, all black hole states in the ensemble are thermalized. Therefore, the stationary solution is asymptotic to
\begin{eqnarray}
\rho(r,v)=Z^{-1} \exp \left\{-\left[\frac{1}{2}v^2+G(x)\right]/T\right\}\;,
\end{eqnarray}
with the normalization constant
\begin{eqnarray}
Z=\frac{2\pi T}{\omega_s} e^{-G(r_s)/T}\;.
\end{eqnarray}
The approximate solution to the stationary Fokker-Planck equation that satisfies the boundary condition $\rho(r>r_m,v)=0$ and is asymptotic to the solution near the bottom of the well is given by \cite{Kampen}
\begin{eqnarray}
\rho(r,v)&=&(2\pi T)^{-3/2} \exp\left[-\frac{G(r)-G(r_s)+\frac{1}{2}v^2}{T}\right]
\frac{\omega_s \omega_m}{\sqrt{\zeta \alpha_+}}\nonumber\\
&&\times \int_{(r-r_m)-\lambda_+ v/\omega_m^2}^{+\infty} \exp\left[-\frac{\omega_m^4 u^2}{2\zeta T \lambda_+}du\right]
\end{eqnarray}
For the intermediate friction coefficient, the MFPT is given by \cite{Kampen}
\begin{eqnarray}\label{inter_fric_formula}
\langle t \rangle&=&k^{-1}=\left( \int_{-\infty}^{+\infty} v\rho(r_m,v)dv \right)^{-1}\nonumber\\&=&\left(\left(1+\frac{\zeta^2}{4\omega_m^2}\right)^{1/2}+\frac{\zeta}{2\omega_m} \right)\frac{2\pi}{\omega_s}\exp\left(\beta W\right)\;.
\end{eqnarray}
This formula is also valid in the large friction regime. For the large friction limit, it reduces to $\langle t \rangle\approxeq \frac{2\pi\zeta}{\omega_{s}\omega_{m}} e^{\beta W}$, which is different from Eq.(\ref{large_fric_formula}) with a factor $2$. This is caused by the different boundary conditions we have used in the derivation.

\section*{Appendix B: Numerical results for different pressures and charges}

We compare the numerical results for different electric charge $Q$ and pressure $P$. Note that the thermodynamic pressure is defined by the relation $P=\frac{3}{8\pi}\frac{1}{L^2}=-\frac{\Lambda}{8\pi}$. Changing the pressure is equivalent to changing the cosmological constant $\Lambda$. The Langevin equation (\ref{LangevinEq}) is solved to collect the trajectories of the state evolution process from the small black hole to the large black hole.

\begin{figure}
  \centering
  \includegraphics[width=8cm]{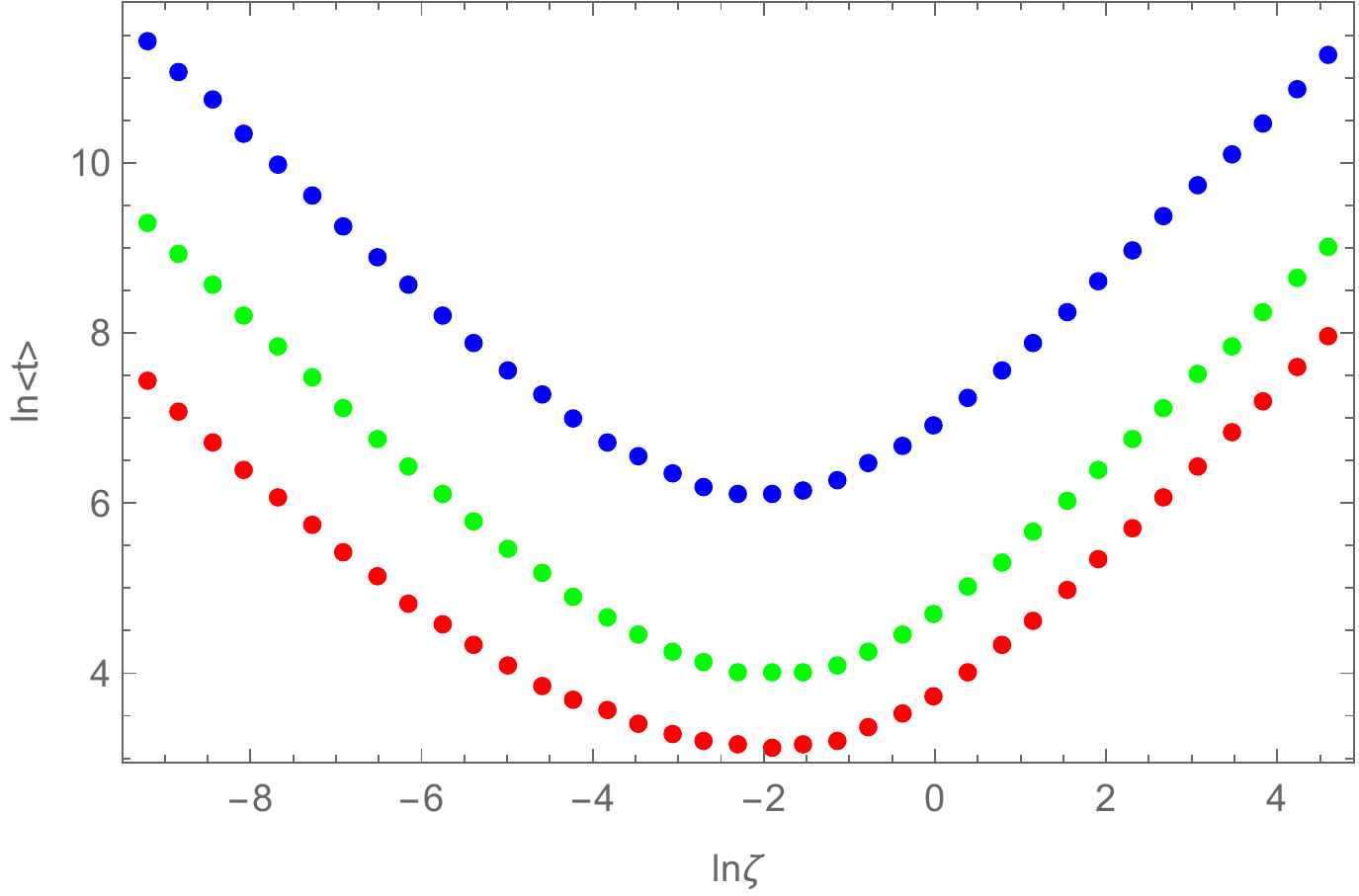}\\
  \caption{The MFPT as the function of the friction coefficient $\zeta$ for different pressure $P$. In this plot, $Q=1$. The pressure $P=0.4Pc, 0.6Pc$ and $0.8Pc$ for the blue, the green, and the red dotted lines. For the given pressure and charge, the phase transition temperature is determined by the equal depths of the double basins as shown in Fig.\ref{GPlot}. The corresponding phase transition temperatures in this plot are $0.0298, 0.0354$  and $0.0396$, respectively. The ratios of the barrier height and the temperature are $3.23,0.841$ and $0.143$.}
  \label{NumResult_diffP_MFPT}
\end{figure}

\begin{figure}
  \centering
  \includegraphics[width=8cm]{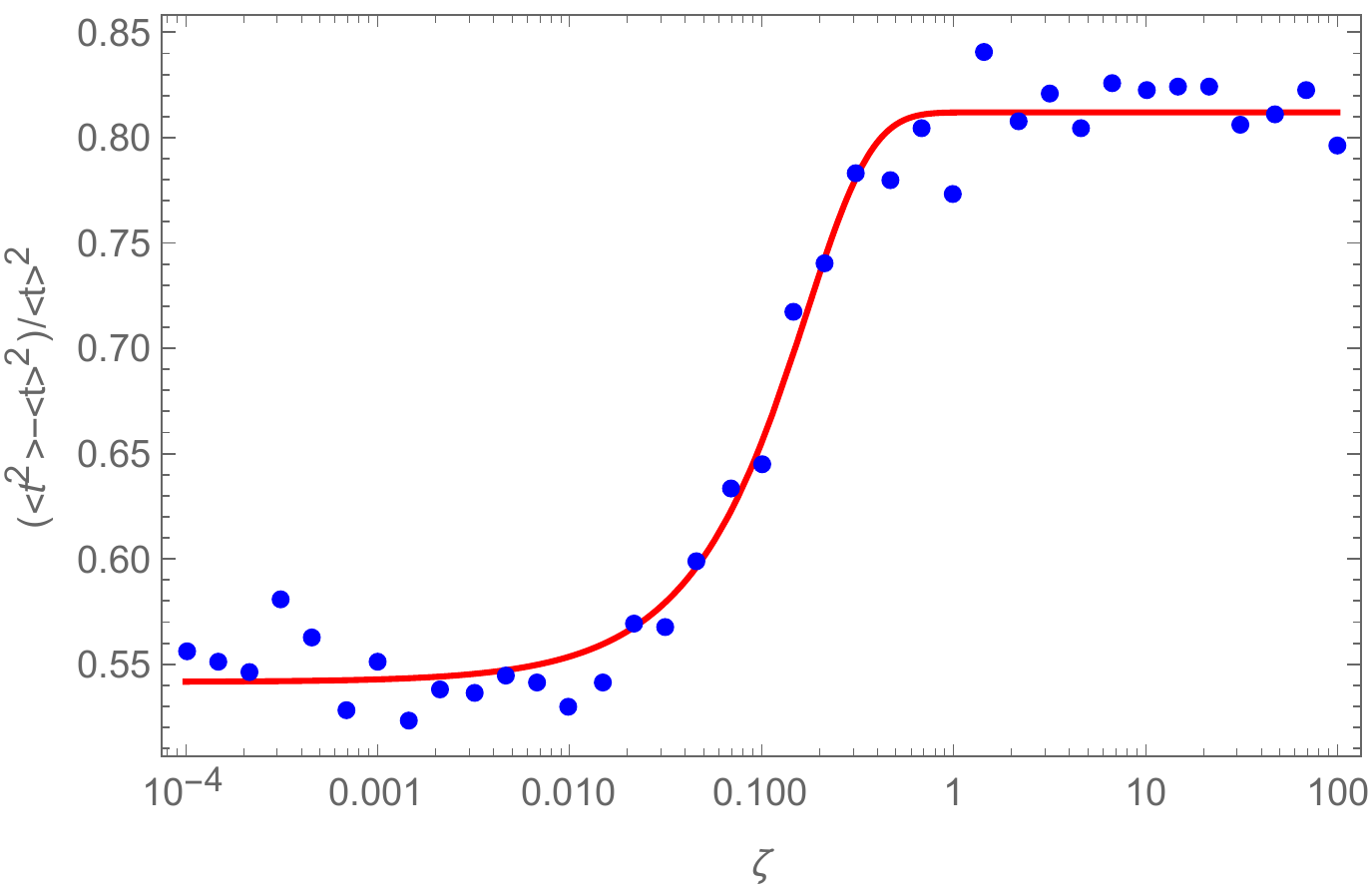}\\
  \includegraphics[width=8cm]{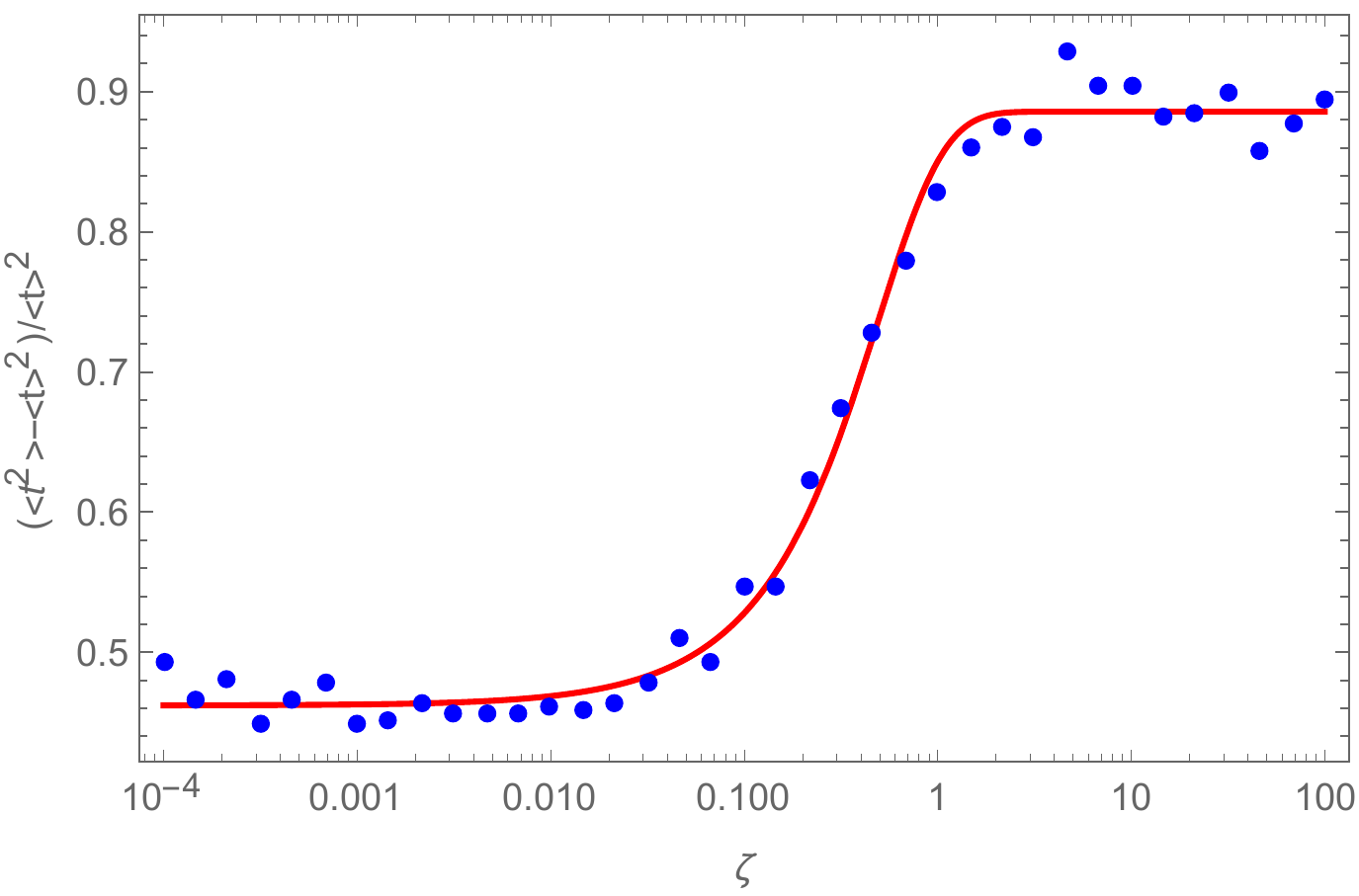}\\
  \caption{The relative fluctuations as the functions of the friction coefficient $\zeta$ for $P=0.6Pc$ (upper panel) and $0.8Pc$ (lower panel). The blue points are the numerical results and the red lines are the fitting functions. The fluctuations for $P=0.4P_c$ are presented in the lower panel of Fig.\ref{NumResult}.}
  \label{NumResult_diffP_Fluc}
\end{figure}

In Fig.\ref{NumResult_diffP_MFPT}, we have plotted the MFPT as the function of the friction coefficient $\zeta$ for different pressure $P$. For $P=0.4P_c$, we have compared the numerical results with the analytical formulas in Fig.\ref{NumResult}. When increasing the pressure, the ratio between the barrier height and the temperature gets smaller. Then the analytical formulas are not valid. In Fig.\ref{NumResult_diffP_MFPT}, we do not compare the analytical formulas with the numerical results. From Fig.\ref{NumResult_diffP_MFPT}, the kinetic turnover can be observed for different pressure. When the pressure increases, the MFPT for the specific friction coefficient decreases. The main reason behind is that the barrier height between the small and the intermediate black holes decreases when increasing the pressure, although the width of the potential well also influences the kinetics \cite{Wei:2021bwy}. The relative fluctuations in kinetics for different pressure are plotted in Fig.\ref{NumResult_diffP_Fluc}. The transitions from the small fluctuations to the large fluctuations are also observed.

\begin{figure}
  \centering
  \includegraphics[width=8cm]{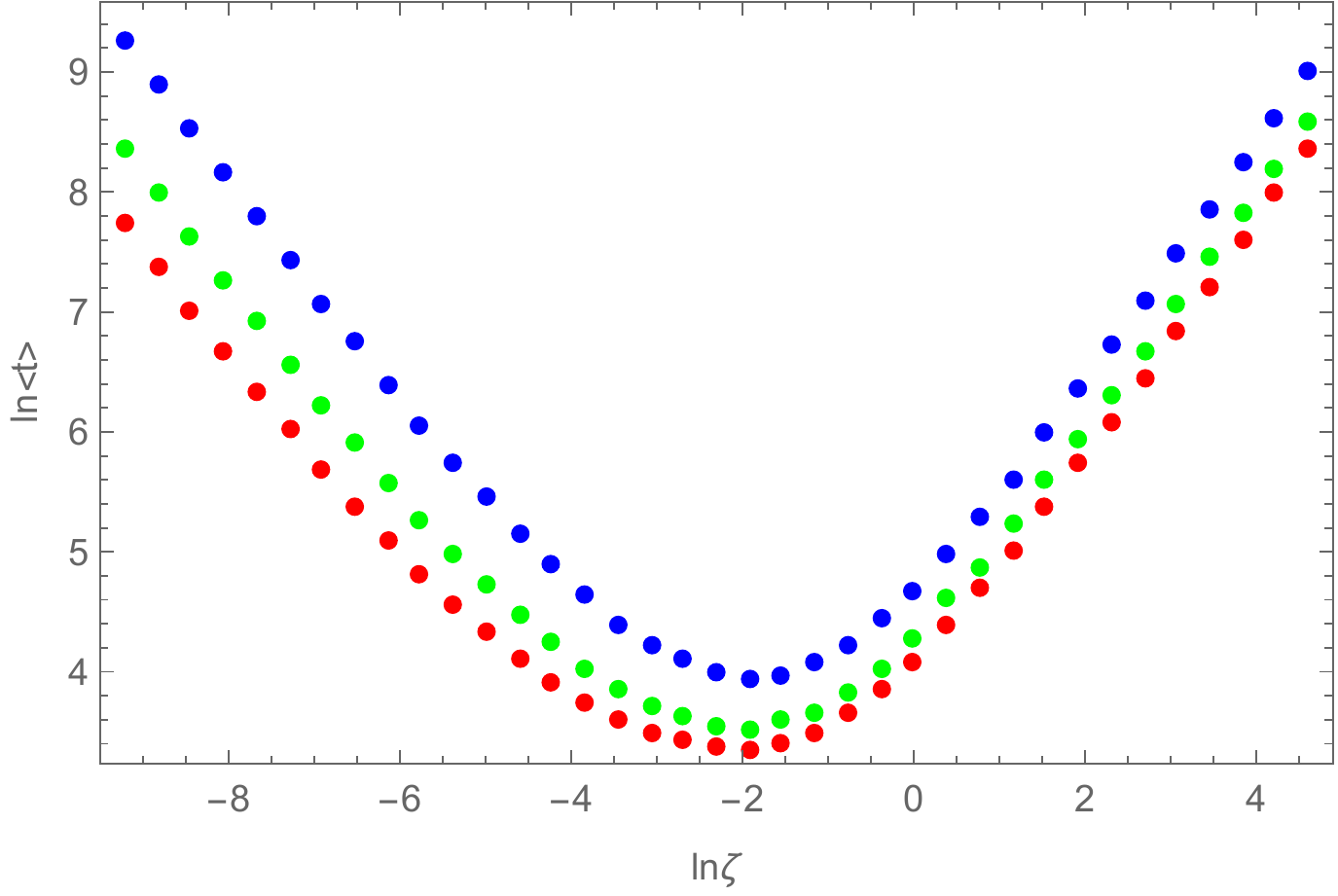}\\
  \caption{The MFPT as the function of the friction coefficient $\zeta$ for different electric charge $Q$. In this plot, $P=0.002$. The charge $Q=1, 1.1$ and $1.15$ for the blue, the green, and the red dotted lines. The corresponding phase transition temperatures are $0.0355, 0.0348$  and $0.0345$. The ratios of the barrier height and the temperature are $0.821,0.357$ and $0.194$.}
  \label{NumResult_diffQ_MFPT}
\end{figure}

\begin{figure}
  \centering
  \includegraphics[width=8cm]{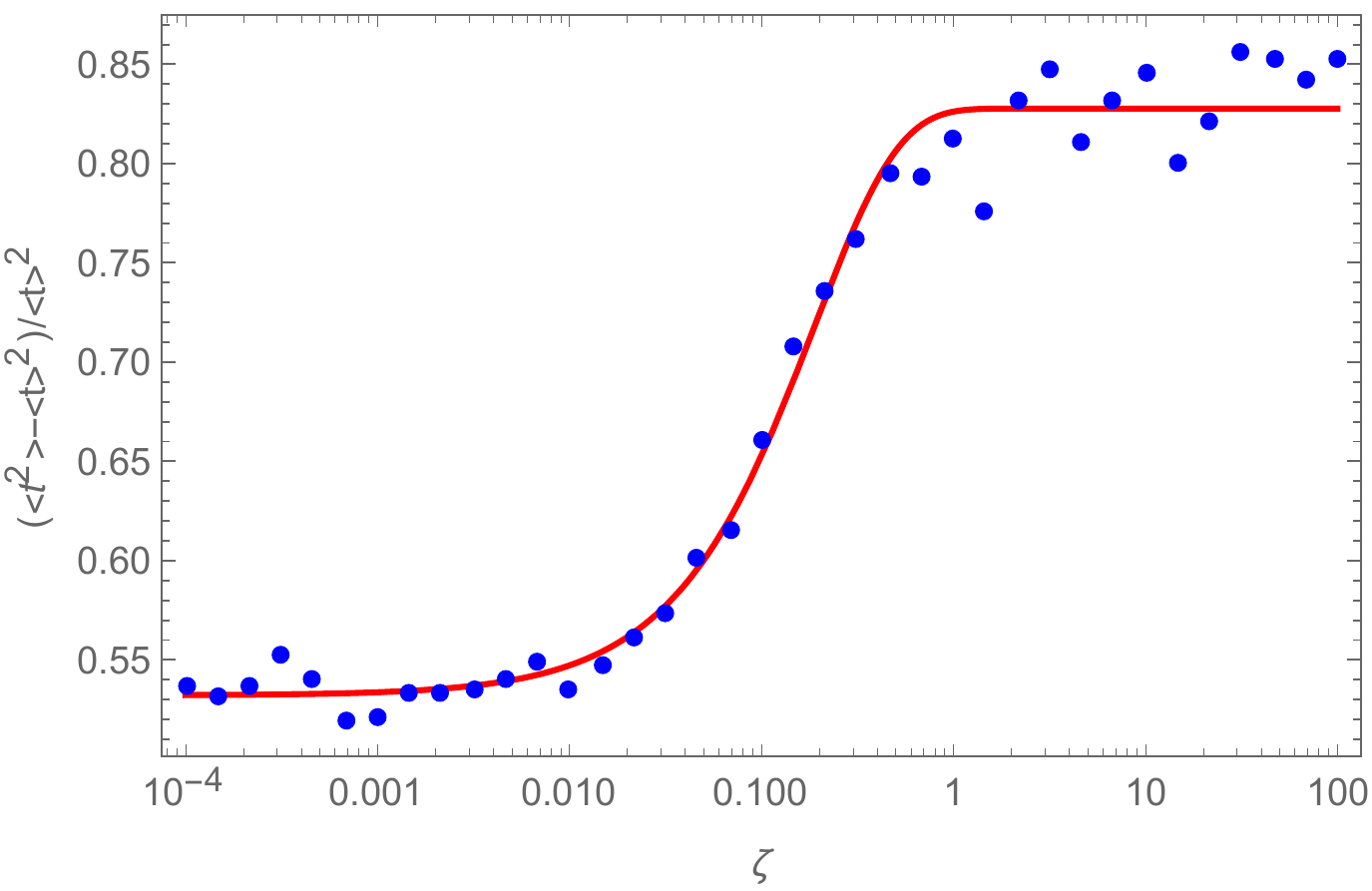}\\
  \includegraphics[width=8cm]{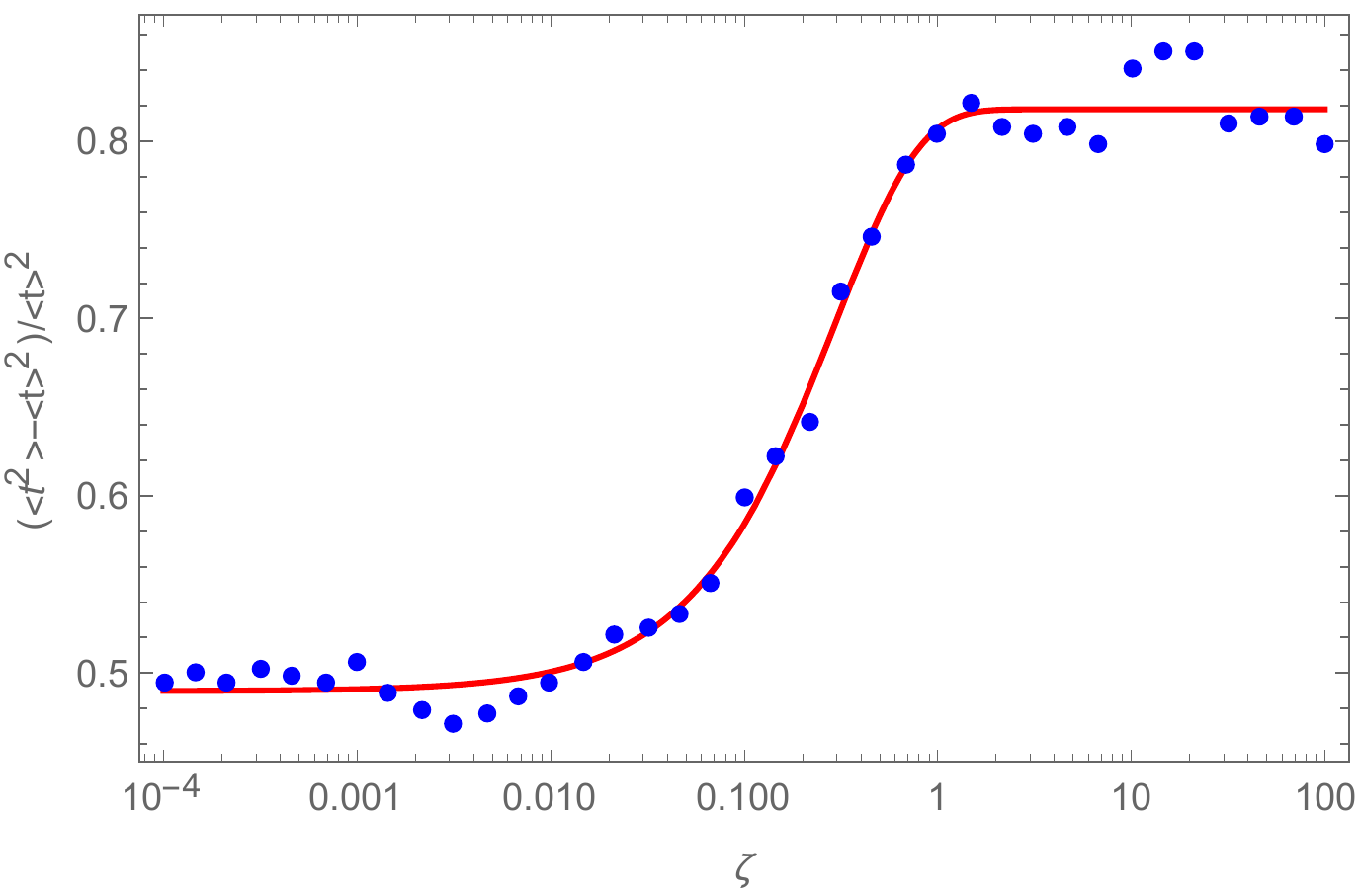}\\
  \includegraphics[width=8cm]{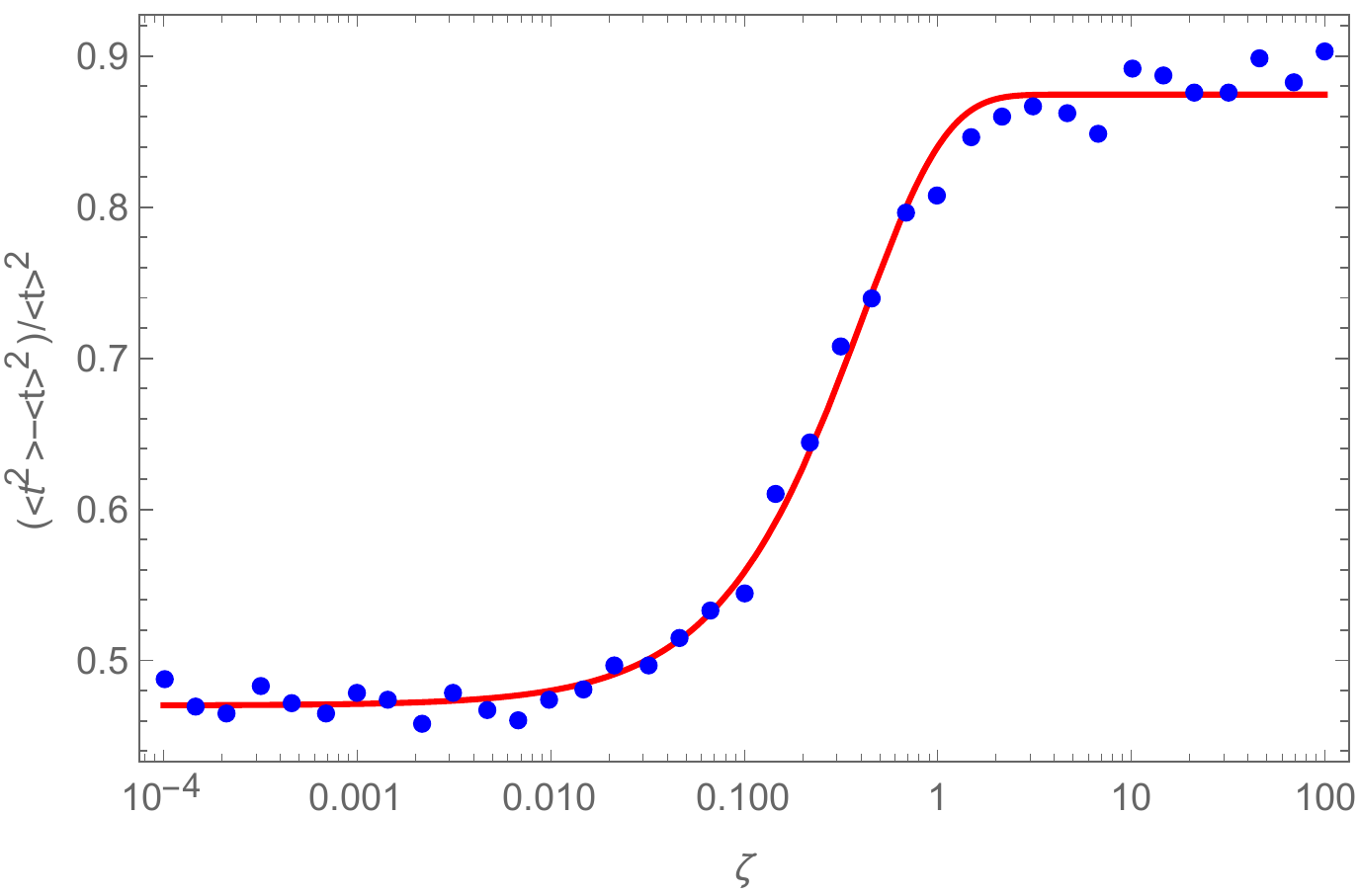}\\
  \caption{The relative fluctuations as the functions of the friction coefficient $\zeta$ for $Q=1$ (upper panel), $Q=1.1$ (middle panel), and $Q=1.15$ (lower panel). The blue points are the numerical results and the red lines are the fitting functions. }
  \label{NumResult_diffQ_Fluc}
\end{figure}

In Fig.\ref{NumResult_diffQ_MFPT}, we have plotted the MFPT as the function of the friction coefficient $\zeta$ for different electric charge $Q$. Because the ratio between the barrier height and the temperature is small, the analytical formulas are also not valid. Only the numerical results are presented. The kinetic turnover is also observed. When the electric charge increases, the MFPT for the specific friction coefficient decreases. This is also caused by the decreasing of the barrier height when increasing the electric charge. The relative fluctuations for different electric charges are plotted in Fig.\ref{NumResult_diffQ_Fluc}. There are transitions from the small fluctuations to the large fluctuations for different charges.

In summary, for different pressure $P$ (cosmological constant $\Lambda$) and electric charge $Q$, we have observed the kinetic turnover and the transition behavior of the relative fluctuations in the phase transition process from the small black hole state to the large black hole through the intermediate black hole state. For the phase transition from the large black hole state to the small black hole state, we expect the same conclusion. This indicates the conclusion for the kinetics of the black hole phase transition is universal. The kinetic turnover and the transition behavior of the relative fluctuations in the black hole phase transition process can then be used to probe the black hole microstructure.

 \end{document}